\documentclass{aa}  
\usepackage{graphicx}
\usepackage{txfonts}

\begin{document}
  \title{Fast 2D non--LTE radiative modelling of prominences}
  \subtitle{I. Numerical methods and benchmark results}

   \author{L. L\'eger\inst{1}
     \and
     L. Chevallier\inst{2,3}
     \and
     F. Paletou\inst{1} }

   \offprints{Ludovick L\'eger}

   \institute{Universit\'e de Toulouse, Observatoire
              Midi-Pyr\'en\'ees, Laboratoire d'Astrophysique de
              Toulouse \& Tarbes (CNRS/UMR 5572), 14 avenue Edouard
              Belin, F-31400 Toulouse, France\\
              \email{lleger@ast.obs-mip.fr, fpaletou@ast.obs-mip.fr}
              \and University of Kentucky, Department of Physics \&
              Astronomy, Lexington, KY 40506-0055, USA\\
              \email{lchevallier@pa.uky.edu}
	      \and Observatoire de Paris-Meudon, LUTh, 5 place Jules Janssen,
              F-92195 Meudon, France}

   \date{Received November 10, 2006 / Accepted March, 2007}

   \abstract
{New high-resolution spectropolarimetric observations of solar
  prominences require improved radiative modelling capabilities in
  order to take into account both multi-dimensional -- at least 2D --
  geometry and complex atomic models.}
{This makes necessary the use of very fast numerical schemes for the
resolution of 2D non--LTE radiative transfer problems considering
freestanding and illuminated slabs.}
{The implementation of Gauss-Seidel and successive over-relaxation
iterative schemes in 2D, together with a multi-grid algorithm, is
thoroughly described in the frame of the short characteristics method
for the computation of the formal solution of the radiative transfer
equation in cartesian geometry.}
{We propose a new test for multidimensional radiative transfer codes
and we also provide original benchmark results for simple 2D
multilevel atom cases which should be helpful for the further
development of such radiative transfer codes, in general.}
{}

   \maketitle

\section{Introduction}

Efficient iterative schemes have been introduced in the field of
two-dimensional (2D) non-LTE numerical radiative transfer during the
last, say, fifteen years. These developments most often rely on the
combination of the short characteristics (SC) method for the so-called
formal solution of the radiative transfer equation (in cartesian
geometry see e.g., Kunasz \& Auer \cite{sc2}, Auer \& Paletou
\cite{lhafp} and in various other geometries, van Noort et
al. \cite{vanoort}) and efficient iterative schemes such as
Gauss-Seidel and successive over-relaxation (GS/SOR) iterative
processes (Trujillo Bueno \& Fabiani Bendicho \cite{tf1}, Paletou \&
L\'eger \cite{paleg}) together with multi-grid (MG) methods (Auer et
al. \cite{lhaft}, Fabiani Bendicho et al. \cite{fta2}).

Hereafter, we are interested in a more realistic modelling of isolated
and illuminated structure, such as prominences hanging in the solar
corona (see e.g., Paletou \cite{fp95}, \cite{fp96}). Our future work
will emphasis on the synthesis of the H and He spectra. Indeed, the
most recent modelling efforts concerning the synthesis of the H and He
spectrum in prominences has been performed using either
mono-dimensional (1D) slabs (Labrosse \& Gouttebroze \cite{lagou1},
\cite{lagou2}), or using 2D cartesian slabs in magnetohydrostatic
equilibrium and the Multilevel Accelerated Lambda Iteration (MALI)
technique for the solution of the non--LTE transfer problem (Heinzel
\& Anzer \cite{phua1}, \cite{phua2}). Concerning the spectrum of H,
Gouttebroze (2006) also presented promising new results using 2D
\emph{cylindrical} models of coronal loops.

Our primary aim is thus to improve diagnostics based, in particular,
on He\,{\sc i} lines by treating a detailed He-atomic model including
the \emph{atomic fine structure} together with 2D non-LTE radiative
transfer. We are motivated here by new solar prominences observations
(see e.g., Paletou et al. \cite{paletal}, Merenda et al. \cite{laura})
which also triggered some revisions of inverting tools (L\'opez Ariste
\& Casini \cite{lopcas}). And up to now, the later diagnostic tools
are limited to the assumption that the relevant (observed) He spectral
lines are optically \emph{thin}; it is, however, easy to check from
high spectral resolution observations that a spectral line like $D_3$
of He\,{\sc i} in the visible, for instance, is not always optically
thin, even in quiescent prominences (Landi Degl'Innocenti
\cite{landi82}, L\'opez Ariste \& Casini \cite{lopcas}). Furthermore,
the expected optical thicknesses of 1 to 10 say, in such structures
let us forecast the presence of significant geometrical effects on the
mechanism of formation of this spectral line, as the ones already put
in evidence on H$\alpha$ by Paletou (\cite{fp97}).

The combination of 2D geometry with a very detailed atomic model for
He obviously requires a more efficient radiative transfer code as
compared to the one developed from the MALI method by Paletou
(1995). And clearly enough, the planned improvement of the radiative
modelling including multidimensional geometry together with
multi-level, realistic atomic models have to rely on those new and
fast radiative transfer methods based on GS/SOR with multigrid
numerical schemes.

GS/SOR methods, best implemented within a short characteristics formal
solver, have been described in every details by Trujillo Bueno \&
Fabiani Bendicho (\cite{tf1}) but \emph{only} in the frame of the
two-level atom case and in 1D geometry. In another article, Fabiani
Bendicho et al. (\cite{fta2}) have nicely described the implementation
of non-linear multi-grid techniques, using an efficient iterative
method such as a GS/SOR scheme. However, they just did \emph{not}
describe the implementation of, 1D or multidimensional, multilevel
GS/SOR scheme using the SC method. Besides, Paletou \& L\'eger
(\cite{paleg}) have finally made \emph{explicit} the implementation of
GS/SOR iterative schemes in the multi-level atom case, restricted
though to a 1D plane-parallel geometry.

The present article aims therefore at ``filling the gap'' by providing
all the elements required for a successful implementation of a GS/SOR
iterative scheme \emph{in a 2D cartesian geometry}. In order to
do so, we adopt the line of detailing the method in the frame of the
2-level atom given that our detailed description of the multilevel
strategy published elsewhere (Paletou \& L\'eger \cite{paleg}) does
not need to be commented any further for the jump from 1D to
2D. Therefore, we also provide hereafter various benchmark results
for the 2D-\emph{multilevel} atom case, still unpublished to date,
using simple atomic models taken from Avrett (\cite{avrett}; see also
Paletou \& L\'eger \cite{paleg}). Moreover, an \emph{original}
comparison between 2D numerical results and independent
\emph{analytical} solutions is made.

We shall recall in \S 2 the basic principles of ALI and GS/SOR
iterative schemes in the frame of a two-level atom model and in 1D
geometry. Then in \S 3 we shall describe, in details, how the GS/SOR
numerical method can be implemented for the case of 2D slabs in
cartesian geometry, therefore upgrading the 2D short characteristic
method initially published by Auer \& Paletou (\cite{lhafp}). A new
test for numerical radiation transfer codes is briefly presented in \S
4. Then we shall finally present, in \S 5, benchmark results for
simplified multilevel atomic models in 2D geometry and some
illustrative examples clearly demonstrating in which conditions
geometrical effects should be seriously considered.

\begin{figure}
  \centering
  \includegraphics[width=8cm,angle=0]{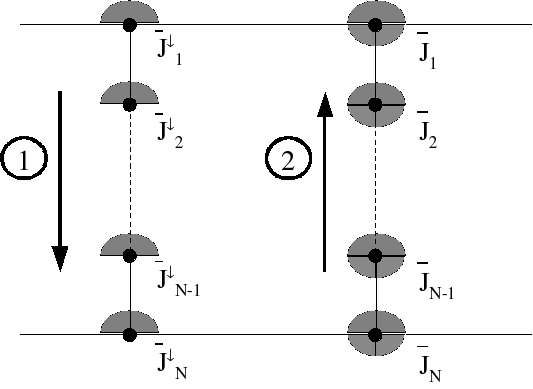}
  \caption{Schematic view of the short characteristics method for the
    computation of $\bar{J}$ in a 1D grid of $N$ points in the case of
    a ALI iterative scheme. During the first (downward here) pass the
    angular integration as in Eq.~(\ref{eq:jbar2}) is made over half
    an hemisphere; it is completed during the 2nd (upward) pass,
    allowing for the source function update according to
    Eq.~(\ref{eq:s2n}).}
  \label{Fig1}
\end{figure}

\section{Gauss-Seidel and SOR iterative schemes basics}

In the \emph{two-level atom case}, the non-LTE line source function, assuming
\emph{complete} redistribution in frequency, is usually written as

\begin{equation}
S(\tau) = (1 - \varepsilon) \bar{J}(\tau) + S^*(\tau) \, ,
\label{eq:s2n}
\end{equation}
where $\tau$ is the optical depth, $S^*$ is the thermal source
function and $\varepsilon$ is the collisional destruction probability;
unless explicitly mentioned, $S^* = \varepsilon B$, where $B$ is the
Planck function. $\bar{J}$ is the usual mean intensity defined as

\begin{equation}
\bar{J} = \oint \displaystyle{{{d \Omega} \over {4 \pi}}}
  \int_{0}^{\infty} {\phi_{\nu} I_{\nu \Omega} d \nu} \, ,
\label{eq:jbar2}
\end{equation}
where the optical depth dependence has been omitted for the sake of
simplicity; as usual, $I_{\nu \Omega}$ is the specific intensity and
$\phi_{\nu}$ is the line absorption profile. Usually again, the mean
intensity is written as the formal solution of the radiative transfer
equation i.e.,

\begin{equation}
\bar{J} = \Lambda [S] \, .
\end{equation}

Following the Jacobi-type iterative scheme introduced in numerical
transfer by Olson et al. (\cite{oab}), we shall consider a splitting
operator $\Lambda^{\ast}$ equal to the exact diagonal of the true
operator $\Lambda$. Now introducing the perturbations

\begin{equation}
\left\{
\begin{array}{lll}
\Lambda & = & \Lambda^{\ast} \, + \, (\Lambda \, - \, \Lambda^{\ast})
\\ S^{\mathrm{(new)}} & = & S^{\mathrm{(old)}} \, + \, \delta S \\
\end{array} \right.
\end{equation}
in Eq.~(\ref{eq:s2n}), we are led to an iterative scheme such that

\begin{equation}
S^{\mathrm{(new)}} = [1 \, - \, (1 \, - \, \varepsilon)
  \Lambda^{\ast}]^{-1} \{ (1 \, - \, \varepsilon)(\Lambda \, - \,
\Lambda^{\ast})S^{\mathrm{(old)}} \, + \, \varepsilon B \} \, .
\end{equation}
Running the later scheme to convergence is better known in numerical
radiative transfer as the ``ALI method''. As schematized in
Fig.~\ref{Fig1}, using the \emph{short characteristics} method in 1D
geometry (Olson \& Kunasz \cite{sc1}, Kunasz \& Auer \cite{sc2}), the
formal solution is obtained by sweeping the grid say, first in
directions $-\Omega$ ($\mu < 0$) i.e., from the surface down to the
bottom of the atmosphere, and then in the opposite, upward directions
$+\Omega$ ($\mu > 0$) starting from the bottom of the atmosphere up to
its surface though. The specific intensity $I_{\nu \Omega}$ is then
advanced step by step during each pass, partially integrated over
angles and frequencies during the downward pass while, during the
second (upward) pass, completion of the angular integration allows for
the full determination of the mean intensity $\bar{J}_{k}$ at each
depth $\tau_{k}$ of the 1D grid and, finally, to the update of the
source function

\begin{equation}
S_{k}^{\mathrm{(new)}} = S_{k}^{\mathrm{(old)}} \, + \, \Delta S_{k}
\, ,
\end{equation}
on the basis of increments such that

\begin{equation}
\Delta S_{k} = { { (1 - \varepsilon) \bar{J}_{k}^{\mathrm{(old)}} +
\varepsilon B_{k} - S_{k}^{\mathrm{(old)}} } \over {1 - (1 -
\varepsilon) \Lambda_{kk} } } \, , \label{eq:incali}
\end{equation}
where $\Lambda_{kk}$ is a \emph{scalar} equal to the diagonal element
of the full operator $\Lambda$ at such a depth in the atmosphere and
where superscripts (old) denote quantities \emph{already known} from the
previous iterative stage.

   \begin{figure}
   \centering
   \includegraphics[width=8cm,angle=0]{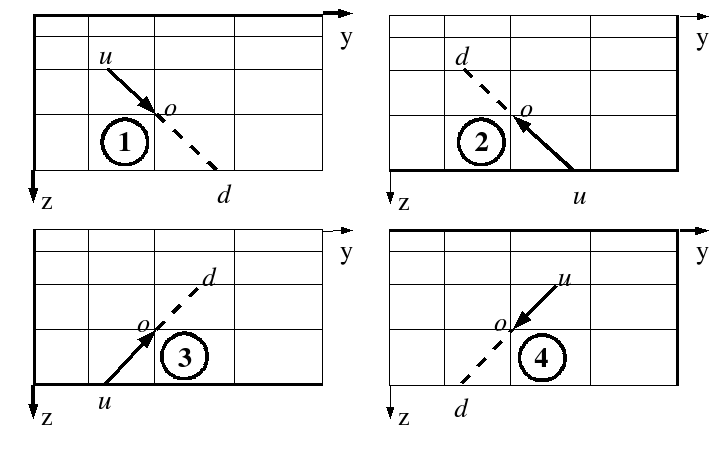}
   \caption{The 2D grid is swept four times: first pass in
     directions $\searrow$ defined in panel 1, second pass in
     directions $\nwarrow$ defined in panel 2, third pass in
     directions $\nearrow$ defined in panel 3 and fourth pass in
     directions $\swarrow$ defined in panel 4. Sweeping must be done
     away from the boundaries so that upwind intensities -- see
      Eq.~(\ref{eq:sc}) -- are always known. }
   \label{Fig2}
   \end{figure}

For a Gauss-Seidel iterative scheme, the sweeping of the atmosphere is
identical \emph{but} as soon as the mean intensity $\bar{J}_{k}$ is
fully computed at depth-point $k$ in the atmosphere during the upward
pass, the \emph{local} source function is updated immediately i.e.,
before completion of the 2nd pass, using increments which have now turned
into

\begin{equation}
\Delta S_{k}^{\mathrm{(GS)}} = { { (1 - \varepsilon)
\bar{J}_{k}^{\mathrm{(old\,and\,new)}} + \varepsilon B_{k} -
S_{k}^{\mathrm{(old)}} } \over {1 - (1 - \varepsilon) \Lambda_{kk} }
} \, , \label{eq:incgs}
\end{equation}
where the quantity $\bar{J}_{k}^{\mathrm{(old\,and\,new)}}$ means that
at the spatial point $k$ the mean intensity has to be calculated via a
formal solution of the transfer equation using the ``new'' source
function values ${S}_{j}^{\mathrm{(new)}}$ already obtained at points
$j= \mathrm{ND}, \ldots, (k+1)$ and the ``old'' source function values
${S}_{j}^{\mathrm{(old)}}$ at points $j=k,\,(k-1), \ldots, 1$.

Finally, as a next step SOR iterations can simply be implemented using

\begin{equation}
\Delta S_{k}^{\mathrm{(SOR)}} = \omega \Delta S_{k}^{\mathrm{(GS)}}
\, , \label{eq:incsor}
\end{equation}
where $\omega$ is an \emph{overrelaxation} parameter such that $1 <
\omega < 2$. For two-level atom models in 1D, this method was
originally proposed by Trujillo Bueno \& Fabiani Bendicho
(\cite{tf1}).

\section{The 2D-cartesian geometry case}

Hereafter, we shall describe \emph{in every details} how the GS/SOR
numerical method can be implemented for the case of 2D \emph{freestanding
slabs} modeled in cartesian geometry.

\subsection{SC in 2D: an overview}

We shall initially follow and therefore \emph{upgrade} the formal
solver of reference proposed originally by Kunasz \& Auer (\cite{sc2})
and modified by Auer \& Paletou (\cite{lhafp}).

Using SC in 2D geometry, the formal solution is obtained by sweeping
the grid four times, as schematized in Fig.~\ref{Fig2}, say first
increasing $y$ and $z$ i.e., along directions $\Omega_{1}$ (note that
$z=0$ is the surface of the atmosphere), second decreasing $y$ and $z$
along directions $\Omega_{2}$, third increasing $y$ and decreasing $z$
along directions $\Omega_{3}$, and finally decreasing $y$ and
increasing $z$ along directions $\Omega_{4}$. The specific intensity
$I_{\nu \Omega}$ is therefore advanced step by step during each pass,
partially integrated over angles, quadrant after quadrant, and over
frequencies during the first three passes while, during the fourth
pass, the mean intensity $\bar{J}$ can be fully computed, completing
therefore the numerical evaluation of the formal solution

\begin{equation}
\bar{J}_{(i,j)} = \Lambda_{(i,j)} [ S ] \, .
\end{equation}

   \begin{figure}
   \centering
   \includegraphics[width=8cm,angle=0]{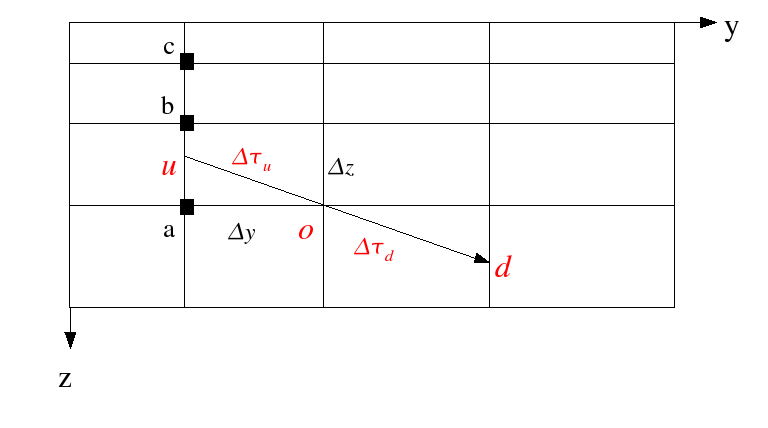}
      \caption{Example of a short characteristic across a 2D cartesian
      grid at depth point $o$ ($i,j$) for a ray propagating along
      directions $\searrow$ defined in panel 1 of Fig.~\ref{Fig2}:
      three points $a$, $b$ and $c$ are used for monotonic parabolic
      interpolation in $z$, in order to evaluate quantities at point
      $u$ following Auer \& Paletou (\cite{lhafp}).  }
         \label{Fig3}
   \end{figure}

Except at the boundary surfaces where the incident radiation is known
a priori, along each direction the specific intensity at the inner
grid points is advanced depth after depth. As displayed in
Fig.~\ref{Fig3}, the short characteristic starts at grid point $o$
($i,j$) and extend in the ``upwind'' and ``downwind'' directions until
it hits one of the cell boundaries either at point $u$ or at point
$d$ that is, \emph{not} grid points in general. The specific intensity
is therefore computed, according to Kunasz \& Auer (\cite{sc2}), as

\begin{equation}
I_{o} = I_{u} e^{- \Delta \tau_{u}} + \Psi_{u} S_{u}+ \Psi_{o} S_{o} +
\Psi_{d} S_{d} \, . 
\label{eq:sc}
\end{equation}
where the first part of the right-hand side of this expression
corresponds to the part transmitted from the ``upwind'' point $u$
down to the current point $o$, and the three last terms result from
the analytic integration of

\begin{equation}
{\cal{I}} = \int_{0}^{\Delta \tau_{u}} {S(\tau) e^{- \tau}} d \tau
\end{equation}
along the short characteristic going from $u$ to $o$; expressions for
the $\Psi$'s can be found in Paletou \& L\'eger (\cite{paleg}).

As shown in Fig.~\ref{Fig3}, in 2D geometry, $I_{u}$, $S_{u}$ and
$S_{d}$ are not grid points, and they must be evaluated by
interpolation on the basis of a set of grid point. In order to do so,
one has first to determine on which axis, $y$ or $z$, the upwind and
downwind points shall lie. We introduce $c_{y}$ (respectively $c_{z}$)
the cosine between the direction into which the photon is moving and
the $y$-axis (respectively the $z$-axis), $\Delta y$ the $y$ length of
the cell containing both $u$ and $o$ grid points, and $\Delta z$ its
length in $z$. If $$\frac{\Delta y}{c_{y}} < \frac{\Delta z}{c_{z}}$$
the ray hits the $y$-axis and $\Delta \tau_{u} = {\Delta z}/{c_{z}}$.

Following Auer \& Paletou (\cite{lhafp}), $I_{u}$ and $S_{u}$ are
determined by interpolation along the upwind grid-line passing through
points $a$ and $b$. To perform a parabolic interpolation, we shall
therefore use three grid points $a$, $b$ and $c$ as displayed in
Fig.~\ref{Fig3}, and where ``quantities'' have already been updated;
along $z$-lines, interpolation weights would be given by

\begin{equation}
\left\{
\begin{array}{l}
\omega_{a} =
\frac{\displaystyle(z_{b}-z)(z_{c}-z)}{\displaystyle(z_{b}-z_{a})(z_{c}-z_{a})}\\\\
\omega_{b} =
\frac{\displaystyle(z_{a}-z)(z_{c}-z)}{\displaystyle(z_{a}-z_{b})(z_{c}-z_{b})}\\\\
\omega_{c} =
\frac{\displaystyle(z_{a}-z)(z_{b}-z)}{\displaystyle(z_{a}-z_{c})(z_{b}-z_{c})}
\end{array}
\right. 
\label{eq:weights}
\end{equation}
and similar weights should be used for interpolation in $y$, using grid
points ($i-1$,$j$), ($i-1$,$j-1$) and ($i-1$,$j-2$) though. Then, we
are able to calculate the upwind specific intensity as

\begin{equation}\label{eq:interp}
I_{u} = \omega_{a} I_{a} + \omega_{b} I_{b} + \omega_{c} I_{c}
\end{equation}
where specific intensity values have already been computed at grid
points $a$, $b$ and $c$. This is guaranteed by sweeping the grid away
from one of the upwind boundaries. Note also that $S_{u}$ and $S_{d}$
are also evaluated from ($S_{a}$, $S_{b}$, $S_{c}$) using similar
expressions.

For the sake of accuracy and in order to avoid the generation of
spurious upwind intensities by high-order interpolation, one must
use a monotonic interpolation i.e., set $I_{u}$ (and $I_{d}$) equal to
the minimum or maximum of $I_{a}$ and $I_{b}$ if the parabolic
interpolant lies outside the interval [min($I_{a}$,$I_{b}$),
max($I_{a}$,$I_{b}$)], as proposed by Auer \& Paletou (\cite{lhafp}).

   \begin{figure}
   \centering
   \includegraphics[width=8cm,angle=0]{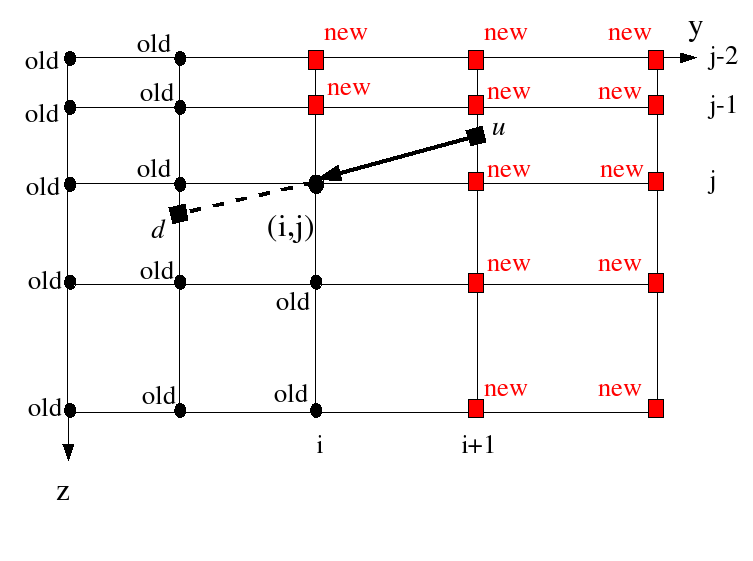}
      \caption{Let $i$ be the index along the $y$-axis and $j$ along
      the $z$-axis; we consider the specific intensity evaluation at any
      inner grid point ($i$,$j$) during the fourth pass in the 2D
      grid, corresponding to the directions $\Omega_4$ defined in
      Fig.~\ref{Fig2}. At this stage, all ``new'' grid
      points have already been swept, so that source functions at
      these points have been updated too.}
         \label{Fig4}
   \end{figure}

\subsection{Implementation of GS/SOR in 2D}

Assume that one has {\em already} swept the grid {\em three times} as
described in Fig.~\ref{Fig2}. By analogy with the GS/SOR numerical
strategy in 1D geometry, we are now going to update the source
function at each grid point during the fourth pass of the SC-2D
scheme, according to the correction given in Eq.~(\ref{eq:incgs})
and \emph{before} passing to the next depth point. It is a quite
straightforward task at the boundary surfaces since the incident
radiation field is known a priori from the (given) external conditions
of illumination.

We shall hereafter describe what has to be done at the inner grid
points. Fig.~\ref{Fig4} describes the situation once arriving at
($i$,$j$) after the 2D grid was swept thrice. Using superscripts
defined in Fig.~\ref{Fig2}, the current specific intensity comes from

\begin{equation}
I_{(i,j)}^{\swarrow} = I_{u}^{\swarrow} e^{-\Delta
\tau_{u}^{\swarrow}} + \Psi_{u}^{\swarrow} S_{u}^{\swarrow
\mathrm{(new)}} + \Psi_{0}^{\swarrow} S_{0}^{\swarrow \mathrm{(old)}}
+ \Psi_{d}^{\swarrow} S_{d}^{\swarrow \mathrm{(old)}} \, ,
\end{equation}
where one must understand quantities with superscripts
$\mathrm{(new)}$ such as resulting from interpolations along upwind
grid lines, using source functions that has been obtained during the
preceding steps. Indeed, using an expression similar to the one in
Eq.~(\ref{eq:interp}), for an interpolation along the $z$-axis, we would have

\begin{equation}
S_{u}^{\swarrow \mathrm{(new)}}=\omega_{a,(i,j)}^{\swarrow}
S_{(i+1,j)}^{\mathrm{(new)}} + \omega_{b,(i,j)}^{\swarrow}
S_{(i+1,j-1)}^{\mathrm{(new)}} + \omega_{c,(i,j)}^{\swarrow}
S_{(i+1,j-2)}^{\mathrm{(new)}} \, .
\end{equation}
Before integrating over all frequencies and over the angles
corresponding to the $\Omega_4$ directions in order to obtain the
partial mean intensity

\begin{equation}
\bar{J}^{\swarrow}=\int_{\Omega_{4}}
\frac{d\Omega}{4\pi} \int_{\nu} {\phi_{\nu} I_{\nu \Omega} d \nu} \, ,
\end{equation}
we shall have to correct the specific intensity calculated during
\emph{the first three passes} for consistency with the source function
updates. More specifically, the term $I_{(i,j)}^{\nearrow}$ was
calculated during the third pass as{\footnote{We use the superscript
(OLD) to emphasize terms which need to be replaced by new values
according to updated points in Fig.~\ref{Fig4} whereas (old) terms
remain unchanged.}}

\begin{equation}
I_{(i,j)}^{\nearrow} = I_{u}^{\nearrow} e^{-\Delta
\tau_{u}^{\nearrow}} + \Psi_{u}^{\nearrow} S_{u}^{\nearrow
\mathrm{(old)}} + \Psi_{0}^{\nearrow} S_{0}^{\nearrow \mathrm{(old)}}
+ \Psi_{d}^{\nearrow} S_{d}^{\nearrow \mathrm{(OLD)}}
\label{dI3}
\end{equation}
using $S_{d}^{\nearrow \mathrm{(OLD)}}$ instead of the \emph{new}
value $S_{d}^{\nearrow \mathrm{(new)}}$ obtained from the
interpolation using the updated points ($i+1$,$j$), ($i+1$,$j-1$) and
($i+1$,$j-2$) as shown in Fig.~\ref{Fig4} since we have the identity
$S_{d}^{\nearrow}=S_{u}^{\swarrow}$ ; a correcting term

\begin{equation}
\Delta \bar{J}_{(i,j)}^{\nearrow} = \int_{\Omega_{3}}
\frac{d\Omega}{4\pi} \int_{\nu} \phi_{\nu} \left[ S_{u,\nu}^{\swarrow
\mathrm{(new)}}-S_{d,\nu}^{\nearrow \mathrm{(old)}} \right] \,
\Psi_{d,\nu}^{\nearrow} d\nu
\label{eq:deltajbar}
\end{equation}
must therefore be added to the the total mean intensity by integrating
the specific intensity correction over frequencies and over all angles
$\Omega_{3}$ -- see Fig.~\ref{Fig2}; this step is equivalent to the
computation of the $\Delta J_{k}^{\mathrm{in}}$ correction mentioned by
Trujillo Bueno \& Fabiani Bendicho (\cite{tf1}) in their Eq.~(39).

\begin{table*}
\caption{Computation time (for a Pentium-4 @ 3 GHz processor) and number of
iterations for the H\,{\sc i} multilevel benchmark model of Avrett
(1968) in a 2D grid with $y_{max}=5\,000$ km and $z_{max}=30\,000$ km
together with 3 angles per octant and 8 frequencies; the temperature
of the atmosphere is 5\,000 K and the gas pressure
$p_{g}\,=\,1\,\mathrm{dyn\,cm^{-2}}$.}
\label{table1}
\centering
\begin{tabular}{c c c c c c}
\hline\hline
Points number & MALI 2D & GSM 2D & SOR 2D & MG 2D & $R_{c}$ \\
\hline
123x123 & 3min9s (46) & 2min19s (29) & 1min17s (16) & 55s (11) & $1.1 \times 10^{-2}$ \\
163x163 & 9min39s (79) & 6min56s (48) & 3min33s (24) & 1min52s (13) & $2.1 \times 10^{-3}$ \\
203x203 & 22min47s (116) & 14min36s (68) & 7min34s (33) & 2min50s (14) & $5.7 \times 10^{-4}$ \\
243x243 & 45min32s (158) & 29min10s (90) & 14min3s (43) & 4min13s (14) & $1.9 \times 10^{-4}$ \\
\hline
\end{tabular}
\end{table*}

The two other terms $I_{(i,j)}^{\searrow}$ and $I_{(i,j)}^{\nwarrow}$
calculated during the first and the second passes are also still
inconsistent with the last source function updates because they were
calculated as

\begin{equation}
\begin{array}{llll}
I_{(i,j)}^{\searrow} & = & & I_{u}^{\searrow \mathrm{(OLD)}}
e^{-\Delta \tau_{u}^{\searrow}} + \Psi_{u}^{\searrow} S_{u}^{\searrow
\mathrm{(OLD)}} \\\\ & & + & \Psi_{0}^{\searrow} S_{0}^{\searrow
\mathrm{(old)}} + \Psi_{d}^{\searrow} S_{d}^{\searrow \mathrm{(OLD)}}
\, ,
\end{array}
\label{dI1}
\end{equation}
and

\begin{equation}
\begin{array}{llll}
I_{(i,j)}^{\nwarrow} & = & & I_{u}^{\nwarrow \mathrm{(OLD)}}
e^{-\Delta \tau_{u}^{\nwarrow}} + \Psi_{u}^{\nwarrow} S_{u}^{\nwarrow
\mathrm{(OLD)}} \\\\ & & + & \Psi_{0}^{\nwarrow} S_{0}^{\nwarrow
\mathrm{(old)}} + \Psi_{d}^{\nwarrow} S_{d}^{\nwarrow \mathrm{(OLD)}}
\, ,
\end{array}
\label{dI2}
\end{equation}
where we have the following identities $S_{u}^{\searrow
\mathrm{(OLD)}}=S_{d}^{\nwarrow \mathrm{(OLD)}}$ and $S_{d}^{\searrow
\mathrm{(OLD)}}=S_{u}^{\nwarrow \mathrm{(OLD)}}$.

These (OLD) source functions could now be calculated using updated
values. For example the \emph{new} value $S_{u}^{\searrow
\mathrm{(new)}}=S_{d}^{\nwarrow \mathrm{(new)}}$ is obtained from an
equation similar to Eq.~(\ref{eq:interp}) with an interpolation along
$y$-axis{\footnote{For an interpolation along $z$-axis, there are no
``new'' grid points to consider.}} using $S_{i,j-1}^{\mathrm{(new)}}$
-- and one can see, using Fig.~\ref{Fig4}, that ($i$,$j-1$) is a ``new'' grid
point whereas ($i-1$,$j-1$) and ($i-2$,$j-1$) are ``old'' grid points i.e.,

\begin{equation}
S_{u}^{\searrow \mathrm{(new)}}=\omega_{a,(i,j)}^{\searrow}
S_{(i,j-1)}^{\mathrm{(new)}} + \omega_{b,(i,j)}^{\searrow}
S_{(i-1,j-1)}^{\mathrm{(old)}} + \omega_{c,(i,j)}^{\searrow}
S_{(i-2,j-1)}^{\mathrm{(old)}}
\end{equation}
Similarly, the \emph{new} value $S_{d}^{\searrow
\mathrm{(new)}}=S_{u}^{\nwarrow \mathrm{(new)}}$ is obtained
using an interpolation along $y$-axis, for instance, involving
$S_{i+1,j+1}^{\mathrm{(new)}}$ and $S_{i+2,j+1}^{\mathrm{(new)}}$ --
with this time, using Fig.~\ref{Fig4}, grid points at ($i+1$,$j+1$)
and ($i+2$,$j+1$) are ``new'' whereas ($i$,$j+1$) is an ``old'' grid
point i.e.,

\begin{equation}
S_{d}^{\searrow \mathrm{(new)}}=\omega_{a,(i,j)}^{\searrow}
S_{(i,j+1)}^{\mathrm{(old)}} + \omega_{b,(i,j)}^{\searrow}
S_{(i+1,j+1)}^{\mathrm{(new)}} + \omega_{c,(i,j)}^{\searrow}
S_{(i+2,j+1)}^{\mathrm{(new)}}
\end{equation}
By analogy, old specific intensities $I_{u}^{\searrow \mathrm{(OLD)}}$
and $I_{u}^{\nwarrow \mathrm{(OLD)}}$ must be updated to obtain
\emph{new} values calculated with interpolations using ``new'' grid
points.

We shall then have to calculate two other corrections $\Delta
J_{(i,j)}^{\searrow}$ and $\Delta J_{(i,j)}^{\nwarrow}$ by integrating
these corrected specific intensities over frequencies and over
directions $\Omega_{1}$ and $\Omega_{2}$, following an equation
similar to Eq.~(\ref{eq:deltajbar}). Finally we shall add three
correcting terms to compute the \emph{correct} total mean intensity at
the current grid point ($i$,$j$):

\begin{equation}
\begin{array}{llll}
\bar{J}_{(i,j)} & = & & \bar{J}_{(i,j)}^{\searrow} \, + \,
\bar{J}_{(i,j)}^{\nwarrow} \, + \, \bar{J}_{(i,j)}^{\nearrow} \, + \,
\bar{J}_{(i,j)}^{\swarrow} \, \\\\ & & + & \, \Delta \bar{J}_{(i,j)}^{\nearrow}
\, + \, \Delta \bar{J}_{(i,j)}^{\searrow} \, + \, \Delta
\bar{J}_{(i,j)}^{\nwarrow}
\end{array} \, .
\end{equation}
Then it is straightforward to update the local source function
$S_{i,j}^{\mathrm{(new)}}$ via Eq.~(\ref{eq:incgs}).

However, before advancing to the next depth point ($i$,$j+1$), it is
important to add the following corrections to the specific intensities
of the three first passes, due to the source function update which has
just been made at the current depth point :

\begin{equation}
\left\{
\begin{array}{l}
\Delta I_{(i,j)}^{\swarrow} = \Psi_{0}^{\swarrow} \left[
S_{(i,j)}^{\mathrm{(new)}} \, - \, S_{(i,j)}^{\mathrm{(old)}}
\right] \\\\
\Delta I_{(i,j)}^{\searrow} = \Psi_{0}^{\searrow} \left[
S_{(i,j)}^{\mathrm{(new)}} \, - \, S_{(i,j)}^{\mathrm{(old)}}
\right] \\\\
\Delta I_{(i,j)}^{\nwarrow} = \Psi_{0}^{\nwarrow} \left[
S_{(i,j)}^{\mathrm{(new)}} \, - \, S_{(i,j)}^{\mathrm{(old)}}
\right]
\end{array}
\right.
\label{eq:deltai2}
\end{equation}
This last stage is analogous to the correction described by Trujillo
Bueno \& Fabiani Bendicho (\cite{tf1}) in their Eq.~(40).

Finally, a two-dimensional SOR iterative scheme is built when, at each
depth-point ($i,j$), the source function is updated according to

\begin{equation}
\Delta S_{(i,j)}^{\mathrm{(SOR)}} = \omega \Delta
S_{(i,j)}^{\mathrm{(GS)}} \, 
\end{equation}
where $\omega$ is computed exactly in the same way as in the 1D case.

\subsection{Additional notes on the whole numerical scheme}

As in the 1D case, implementing a GS/SOR solver requires to properly
order the various loops; starting from outer to inner loop one may
find: (1) the directions $\Omega_{i}$ as shown on Fig.~\ref{Fig2}, (2)
the direction cosines in each quadrant $\Omega_{i}$ and, finally (3)
the frequencies. The corrections described in Eqs.~(\ref{dI3}),
(\ref{dI1}) and (\ref{dI2}) require some bookkeeping of variables
such as all the $\Psi_{u}$'s and the $\Psi_{d}$'s computed during the
three first passes (for the further computation of the mean
intensity).

Details upon the implementation of GS/SOR for \emph{multilevel} atom
models were given by Paletou \& L\'eger (\cite{paleg}). The main
difference with the two-level atom case is the propagation of the
effects of the local population update: it generates \emph{for each
allowed transition} changes in the absorption coefficients at line
center \emph{and} in the line source functions.

Furthermore, we have also embedded the above-described 2D-GS/SOR
scheme into a nested \emph{multigrid} radiative transfer method
following the precise description given by Fabiani Bendicho et
al. (\cite{fta2}). We use three grids with a grid-doubling
strategy. On the \emph{coarsest} grid (i.e., level $l=1$), we iterate
to convergence i.e., until $R_{c}$ i.e., the relative error on the
level-populations from an iteration to another is ``small'' using the
2D-GS/SOR scheme. For each grid $l=2,3$ where grid level $l=3$ is the
finest one, we interpolate populations onto grid level $l$ using
those obtained onto grid level $(l-1)$ and calculate the corresponding
absorption coefficients and source functions. We iterate onto grid
level $l$ using the \emph{standard multigrid method} from grid level
$l$ down to grid level $l=1$ only until the following stopping
criterion is satisfied

\begin{equation}
R_{c}(iter,l)\frac{\lambda}{1-\lambda} < \frac{1}{8} R_{c}(iter=1,l)
\label{eq:stop}
\end{equation}
where $\lambda \simeq R_{c}(iter,l)/R_{c}(iter-1,l)$, as proposed by Auer
et al. (\cite{lhaft}).

We remind here the main steps of one standard multigrid iteration:
 make one pre-smoothing iteration onto grid level $l$ using a pure
GS iterative scheme, then a restriction down to grid level
$l=1$ to compute the coarse-grid equation, solve the coarse-grid
equation onto grid level $l=1$ using the 2D-SOR scheme, make a
prolongation up to grid level $l$ to obtain a new estimate of the
populations, then one post-smoothing iteration onto grid level $l$
using again a pure GS iterative scheme (it is important to
note that one must make one pre- or post-smoothing iteration on each
grid level using a pure GS iterative scheme). We used a
cubic-centered interpolation for the prolongation and the adjoint of a
nine-point prolongation for the restriction (see e.g., Hackbusch
\cite{hackbusch}).

\section{Validation vs. an analytical solution}

There is \emph{no} analytical solution for 2D non--LTE radiative
transfer. However it is possible to compare 2D numerical solutions to
1D solutions for which accurate and robust numerical and analytical
methods exist. In order for this comparison to be accurate, the slab
has to be sufficiently extended in the $y$ direction
i.e. ``effectively'' infinite.

We have used the ARTY code for the computation of reference,
analytical solutions (Chevallier \& Rutily 2005; see also the
Appendix) obtained using the method of the finite Laplace transform.
This code can solve, indeed, standard 1D problems with an intrinsic
accuracy better than $10^{-10}$; it has already been useful in order
to test the ALI method plus a SC formal solver in 1D for the case of a
non-illuminated, homogeneous and isotropic plane-parallel slab with
internal, homogeneous sources (Chevallier et al. 2003).

   \begin{figure}
   \centering
   \includegraphics[width=8cm,angle=0]{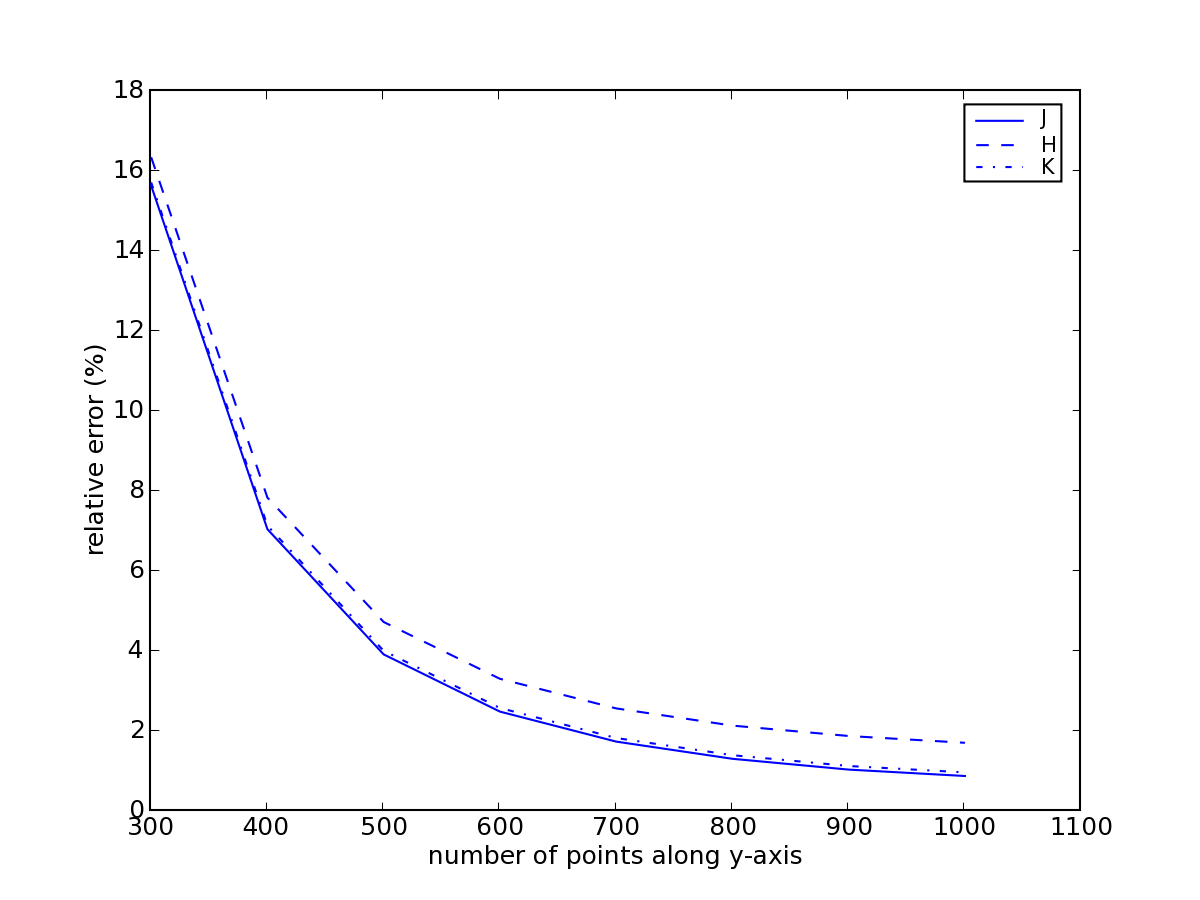}
      \caption{Relative errors between spatial averages of the
      angular moments $J$, $H$ and $K$ given by the 2D-SOR 2-level
      iterative process and their analytical values for
      $\epsilon=0.01$ vs. the number of spatial points of a square 2D
      grid of extension $\tau^*=100$ in each direction.}
         \label{Fig5}
   \end{figure}

A stringent test for our 2D code was to consider a point-source
located at the center of a non-externally illuminated and homogeneous
slab. The central source emits isotropically in space, which in fact
corresponds to a line, infinite along $x$, of sources. This idealized
model captures most of the difficulties met by the numerical methods
to solve the radiative transfer equation: the scattering is not
neglected -- it can also be dominant --, and the ponctual source will
lead to large gradients difficult to handle when dealing with discretization
of the slab.

Then we have computed the properties of the radiation field emerging
at the top surface of the slab ($z=0$) at one frequency, as described
by the usual first three moments $J$, $H$, $K$ of the specific
intensity; more precisely, the later were integrated in space,
\emph{along the top surface of the slab}, in order to be compared to
the 1D analytical solutions.

To achieve this test, we have chosen the difficult case of a slab of
optical thickness $\tau^* = 100$ in both directions where scattering
dominates the absorption adopting the value $\varepsilon=0.01$. This
medium is therefore effectively thick because the thermalization depth
${\ell} \approx 1/\sqrt{3\varepsilon}$ is much less than the optical
thickness in such as case.  We used Carlson's ``Set A''
(\cite{carlson}) with 10 points per octant to describe the angular
dependence of the radiation field and only one frequency-point. The
Dirac thermal source term has been modelized by a sharp, normalised
2D-Gaussian function having half-width at half-maximum 0.16 in $\tau$.
The grid is logarithmically refined near the center of the slab in
order to describe accurately the shape of the 2D-Gaussian: the closest
points from the center are at a distance $10^{-4}$ along the axes, in
order to accurately describe the gaussian shape whose numerical
integral over the space has to be the closest as possible to unity.

The two-level 2D-SOR iterative process was iterated until convergence
of $J$, $H$ and $K$ i.e., when the second digit of their relative
error did not show any more variation from one iteration to the other.
For such a case, 500 iterations are sufficient. In Fig.~\ref{Fig5}, we
demonstrate how these errors behave with the refinement of the spatial
quadratures; the absolute values of the reference solutions are given
in Appendix, as well as the source functions and values of the
specific intensity in the directions corresponding to the angular
quadrature chosen here.

\begin{figure}
\centering
\includegraphics[width=8cm,angle=0]{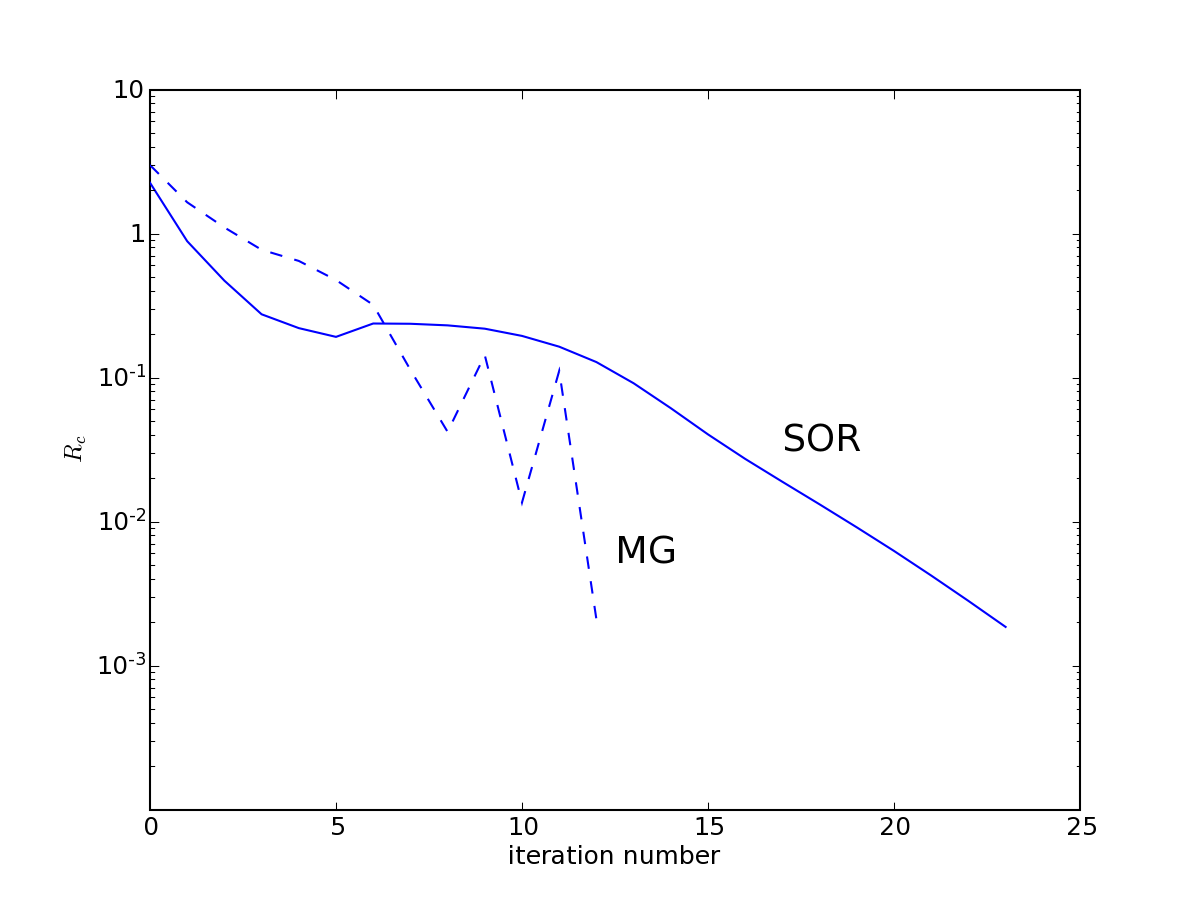}
\caption{Rates of convergence for the SOR (solid lines) and
SOR-MG (dashed lines) 2D multilevel iterative processes. A
spatial grid of 163 points per direction with $z_{max}=30\,000$
km and $y_{max}=5\,000$ km; the temperature of the atmosphere is
5\,000 K and the gas pressure fixed at
$p_{g}\,=\,1\,\mathrm{dyn\,cm^{-2}}$. For benchmark purposes, we
adopted the simplified 3-level H\,{\sc i} atom model taken from
Avrett (\cite{avrett}). }
\label{Fig6}
\end{figure}

The important point to rise here concerning this new test is that (i)
acceptable relative errors, say better than 5\% are obtained only for
very refined grid which (ii) can hardly be handled using a simple
Jacobi-like iterative scheme such as ALI. This justify again the
adoption of very high rate of convergence methods such as GS/SOR plus
MG. Finally, we are conducting more comprehensive tests of this nature
which results will be published elsewhere.


\section{Illustrative examples and benchmarks}

We modeled a 2D freestanding slab irradiated from below on its sides
and bottom by a Planck function. The slab is homogeneous and static
with a vertical \emph{geometrical} extension $z_{max}=30\,000$ km; its
horizontal extension $y_{max}$ could take the respective values:
100\,000, 30\,000, 10\,000, 5\,000 and 1\,000 km. Depth points are
logarithmically spaced away from the boundary surfaces and the
graphical representation we adopted compresses the central region and
\emph{greatly} expand the areas near the boundaries. We have used the
``set A'' of Carlson (\cite{carlson}) with 3 points per octant to
describe the angular dependence of the radiation field and constant
Doppler profiles. The temperature of the slab was fixed to $T$=5\,000
K and the gas pressure $p_{g}=1$~ dyn\,cm$^{-2}$. Finally, we adopted
the standard benchmark models for multilevel atom problems proposed by
Avrett (\cite{avrett}; see also Paletou \& L\'eger \cite{paleg})
considering, in particular, its 3-level H\,{\sc i} atomic model.

   \begin{figure}
   \centering
   \includegraphics[width=8cm,angle=0]{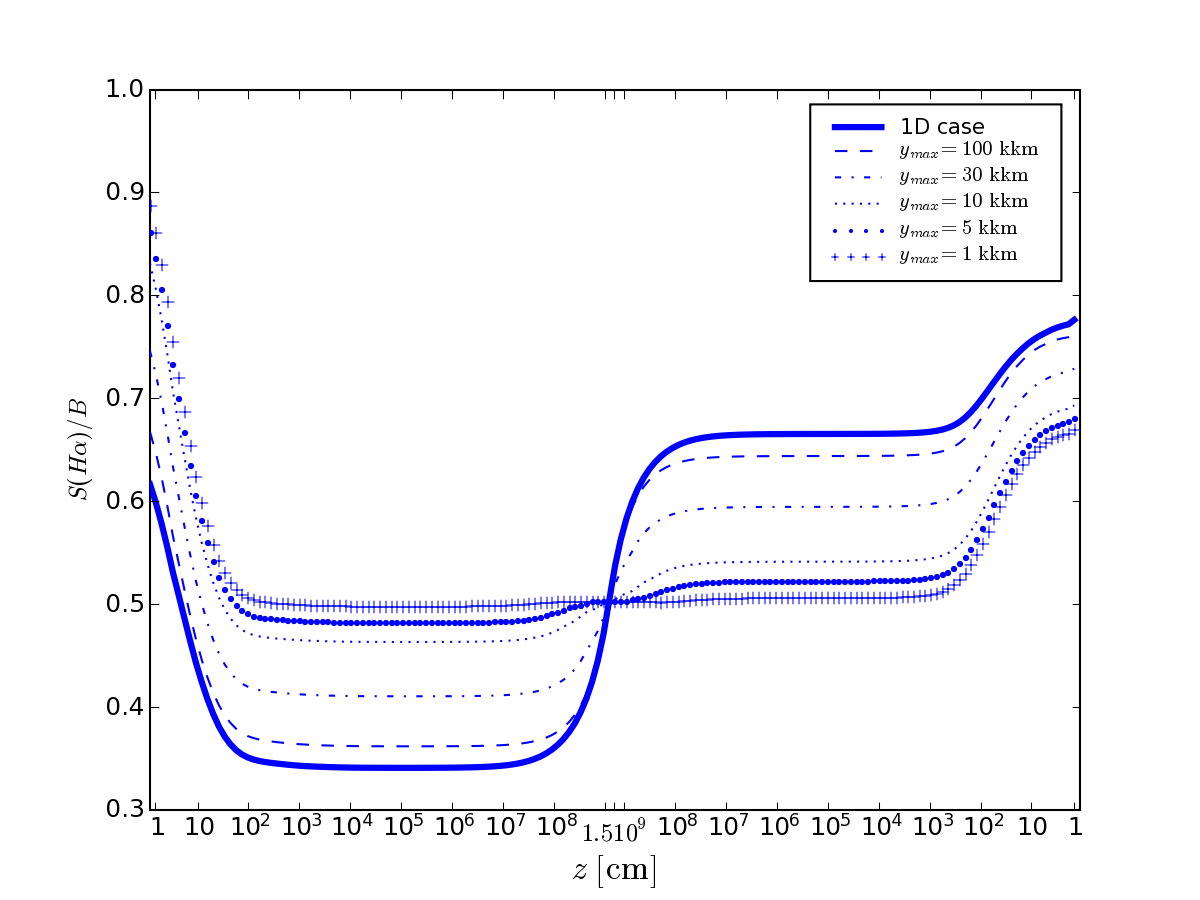}
\caption{Vertical variations of the H$\alpha$ source function (in
      units of B) along the symmetry axis of the 2D slab with
      $z_{max}\, =\, 30\,000$ km and different horizontal extensions
      $y_{max}$ ranging from 1\,000 km to $\infty$ (1D); the
      temperature of the atmosphere is 5\,000 K and the gas pressure
      $p_{g}=1\, \mathrm{dyn\, cm^{-2}}$. The solid line represents
      S(H$\alpha$)/B variations computed in 1D. Note that abscissae give
      \emph{geometrical} positions computed downwards from the top
      surface up to mid-slab and then, symmetrically, upward from slab
      bottom.}
\label{Fig7}
   \end{figure}

The respective rates of convergence for the SOR and MG-2D multilevel
iterative processes are displayed in Fig.~\ref{Fig6} where we have
plotted the maximum relative change on the level populations (i.e.,
the $\infty$-norm) from an iteration to another $R_{c}$. The
computation time for the MALI, GS, SOR and MG 2D-multilevel iterative
processes are given in Tab.~\ref{table1} for different grid
refinements. We remind that a MG scheme is not only superior in
iteration numbers and computing time: it is also important to note
that the convergence error $C_{e}$ which is defined by

\begin{equation}
C_{e} = \mathrm{max} \left( { {\mid n(itr) - n(\infty) \mid} \over
{n(\infty)} } \right)
\end{equation}
is smaller than $R_{c}$ for MG whereas for methods such as MALI or SOR
a small value of $R_{c}$ does not imply a small value of $C_{e}$,
which means that convergence is not necessarily achieved (Fabiani
Bendicho et al. \cite{fta2}).

As shown in Fig.~\ref{Fig7}, where S(H$\alpha$) normalized to the
external illumination is plotted as a function of the vertical
line-center optical depth, the same variations as in 1D (solid line)
are recovered along the vertical axis of symmetry of the 2D model
which has the largest horizontal extension (i.e., 10\,0000 km). For
smaller geometrical slab widths and accordingly horizontal optical
thickness, lateral radiative transfer effects take place and
progressively affect the excitation within the slab. Note that for the
smallest width (i.e., 1\,000 km), we properly recover an almost
constant value $S/B=0.5$ consistent with optically thin conditions
along the horizontal extension of the 2D slab. As first reported by
Paletou (\cite{fp97}), we also recover here under which conditions 2D
radiative transfer effects on the H$\alpha$ source function vertical
variations can be significant; more generally, such effects are a
priori expected for any other spectral lines of moderate optical
thickness.

   \begin{figure}
   \centering
   \includegraphics[width=8cm,height=12cm,angle=0]{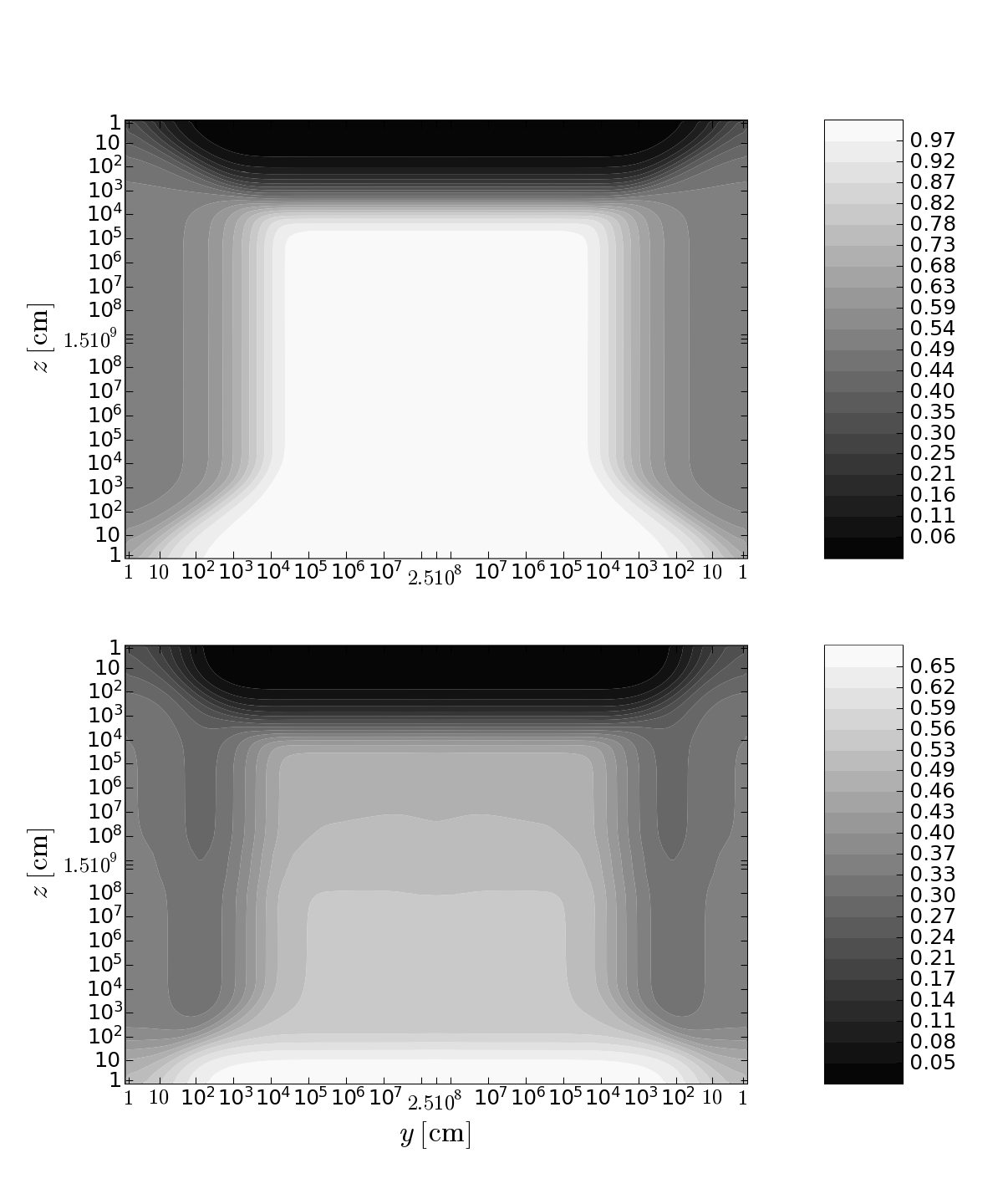}
      \caption{Contour diagram of (top) the second-level populations
      $n_{2}/n_{2}^{\ast}$ and (bottom) the third-level populations
      $n_{3}/n_{3}^{\ast}$ in a grid with $y_{max}$=5 kkm,
      $z_{max}$ =30 kkm, T=5\,000 K and $p_{g}=1\, \mathrm{dyn\,
      cm^{-2}}$. The slab is illuminated from below with a Planck
      function and $n_{2}^{\ast}$ and $n_{3}^{\ast}$ are LTE
      values. Horizontal and vertical axes are defined as described in
      Fig.~\ref{Fig7}. }
         \label{Fig8}
   \end{figure}

Figures~\ref{Fig8} are contour plots of the two excited levels of
hydrogen normalized to their LTE values obtained for a 30\,000 km by
5\,000 km slab. They show both departures from LTE together with
geometry effects within the slab atmosphere. However, since such data
do not exist yet in the litterature while needs for multidimensional
radiative modelling tools are more and more obvious, and in order to
detail the information content of Fig.~\ref{Fig8} about the
populations distribution across the 2D slab, we found that
Tab.~\ref{table2} can also be highly valuable for benchmark purposes.

\begin{table*}
\caption{Second-level (top) and third-level (bottom) populations for
the Avrett (\cite{avrett}) H\,{\sc i} atomic model in a 2D grid of 163
points per direction with $y_{max}\, =\, 5000\, \mathrm{km}$ and
$z_{max}\, =\, 30000\, \mathrm{km}$ together with 3 angles per octant
and 8 frequencies ; the temperature of the atmosphere is 5000 K and
the gas pressure $p_{g}\, =\, 1\, \mathrm{dyn\, cm^{-2}}$.}
\label{table2}
\centering
\begin{tabular}{c c c c c c c c c c c}
\hline \hline
 & & & & & y-position & & & & &\\
                    z-position
 &$1$                           
 &$10$                          
 &$10^{2}$                      
 &$10^{3}$                      
 &$10^{4}$                      
 &$10^{5}$                      
 &$10^{6}$                      
 &$10^{7}$                      
 &$10^{8}$                      
 &$2.5 \times 10^{8}$           
 \\
\hline
$1$                           
  &  93.6107                &  52.4326                &  15.8861                &  6.29160    
  &  5.01322                &  5.02310                &  5.02508                &  5.02511    
  &  5.02517                &  5.02521    
 \\
$10$                          
  &  113.845                &  84.3495                &  29.7678                &  11.8603    
  &  9.45411                &  9.47290                &  9.47663                &  9.47669    
  &  9.47682                &  9.47690    
 \\
$10^{2}$                      
  &  136.951                &  128.456                &  86.6607                &  38.9345    
  &  31.1428                &  31.2092                &  31.2215                &  31.2217    
  &  31.2222                &  31.2225    
 \\
$10^{3}$                      
  &  145.777                &  145.248                &  141.607                &  121.020    
  &  104.541                &  104.935                &  104.976                &  104.977    
  &  104.979                &  104.980    
 \\
$10^{4}$                      
  &  148.677                &  150.732                &  159.910                &  188.041    
  &  229.865                &  237.425                &  237.537                &  237.539    
  &  237.543                &  237.545    
 \\
$10^{5}$                      
  &  149.219                &  151.755                &  163.283                &  199.744    
  &  265.198                &  290.210                &  291.726                &  291.727    
  &  291.732                &  291.736    
 \\
$10^{6}$                      
  &  149.257                &  151.825                &  163.517                &  200.529    
  &  267.018                &  294.048                &  296.308                &  296.366    
  &  296.371                &  296.375    
 \\
$10^{7}$                      
  &  149.259                &  151.829                &  163.530                &  200.576    
  &  267.127                &  294.182                &  296.502                &  296.640    
  &  296.645                &  296.648    
 \\
$10^{8}$                      
  &  149.255                &  151.823                &  163.517                &  200.557    
  &  267.101                &  294.153                &  296.473                &  296.611    
  &  296.597                &  296.593    
 \\
$10^{9}$                      
  &  149.243                &  151.801                &  163.469                &  200.491    
  &  267.010                &  294.051                &  296.370                &  296.509    
  &  296.499                &  296.494    
 \\
$1.5 \times 10^{9}$           
  &  149.241                &  151.798                &  163.462                &  200.481    
  &  266.996                &  294.036                &  296.355                &  296.495    
  &  296.495                &  296.494    
 \\
$3 \times 10^{9} - 10^{9}$    
  &  149.239                &  151.795                &  163.455                &  200.471    
  &  266.983                &  294.020                &  296.339                &  296.480    
  &  296.489                &  296.493    
 \\
$3 \times 10^{9} - 10^{8}$    
  &  149.227                &  151.773                &  163.408                &  200.404    
  &  266.891                &  293.918                &  296.236                &  296.378    
  &  296.392                &  296.396    
 \\
$3 \times 10^{9} - 10^{7}$    
  &  149.223                &  151.767                &  163.394                &  200.386    
  &  266.866                &  293.890                &  296.207                &  296.348    
  &  296.344                &  296.341    
 \\
$3 \times 10^{9} - 10^{6}$    
  &  149.223                &  151.766                &  163.393                &  200.384    
  &  266.863                &  293.886                &  296.204                &  296.343    
  &  296.338                &  296.334    
 \\
$3 \times 10^{9} - 10^{5}$    
  &  149.223                &  151.767                &  163.396                &  200.395    
  &  266.887                &  293.944                &  296.201                &  296.337    
  &  296.331                &  296.328    
 \\
$3 \times 10^{9} - 10^{4}$    
  &  149.286                &  151.886                &  163.789                &  201.752    
  &  270.398                &  294.453                &  296.151                &  296.261    
  &  296.256                &  296.253    
 \\
$3 \times 10^{9} - 10^{3}$    
  &  150.529                &  154.236                &  171.626                &  228.713    
  &  285.362                &  295.302                &  296.043                &  296.091    
  &  296.086                &  296.083    
 \\
$3 \times 10^{9} - 10^{2}$    
  &  157.815                &  168.092                &  216.220                &  274.214    
  &  292.499                &  295.409                &  295.629                &  295.643    
  &  295.639                &  295.636    
 \\
$3 \times 10^{9} - 10$        
  &  179.929                &  210.210                &  266.745                &  287.561    
  &  293.106                &  293.988                &  294.055                &  294.059    
  &  294.057                &  294.055    
 \\
$3 \times 10^{9} - 1$         
  &  199.795                &  241.337                &  278.860                &  289.983    
  &  292.923                &  293.390                &  293.426                &  293.428    
  &  293.427                &  293.426    
 \\
\hline \hline
 & & & & & y-position & & & & &\\
z-position & $1$ & $10$ & $10^{2}$ & $10^{3}$ & $10^{4}$ & $10^{5}$ & $10^{6}$ & $10^{7}$ & $10^{8}$ & $2.5 \times 10^{8}$\\
\hline
$1$                           
  &  1.86186                &  1.39630                & 0.486537                & 0.153291    
  & 0.118500                & 0.118255                & 0.118259                & 0.118240    
  & 0.118095                & 0.118004    
 \\
$10$                          
  &  2.04988                &  1.69002                & 0.627277                & 0.207928    
  & 0.161807                & 0.161613                & 0.161631                & 0.161604    
  & 0.161402                & 0.161275    
 \\
$10^{2}$                      
  &  2.39169                &  2.20327                &  1.31802                & 0.534052    
  & 0.422122                & 0.422372                & 0.422476                & 0.422406    
  & 0.421855                & 0.421507    
 \\
$10^{3}$                      
  &  2.59030                &  2.47736                &  2.04905                &  1.61663    
  &  1.38949                &  1.39398                &  1.39445                &  1.39422    
  &  1.39235                &  1.39117    
 \\
$10^{4}$                      
  &  2.65493                &  2.56695                &  2.29196                &  2.50136    
  &  3.04375                &  3.14272                &  3.14409                &  3.14357    
  &  3.13933                &  3.13665    
 \\
$10^{5}$                      
  &  2.66696                &  2.58361                &  2.33664                &  2.65572    
  &  3.50992                &  3.83926                &  3.85912                &  3.85848    
  &  3.85327                &  3.84998    
 \\
$10^{6}$                      
  &  2.66789                &  2.58487                &  2.33988                &  2.66628    
  &  3.53420                &  3.89018                &  3.92008                &  3.92026    
  &  3.91497                &  3.91163    
 \\
$10^{7}$                      
  &  2.66898                &  2.58607                &  2.34158                &  2.66891    
  &  3.53832                &  3.89491                &  3.92572                &  3.92885    
  &  3.92452                &  3.92119    
 \\
$10^{8}$                      
  &  2.67909                &  2.59711                &  2.35631                &  2.68839    
  &  3.56437                &  3.92360                &  3.95464                &  3.95879    
  &  3.97321                &  3.97675    
 \\
$10^{9}$                      
  &  2.71539                &  2.63671                &  2.40913                &  2.75824    
  &  3.65778                &  4.02648                &  4.05820                &  4.06111    
  &  4.07079                &  4.07557    
 \\
$1.5 \times 10^{9}$           
  &  2.72068                &  2.64248                &  2.41682                &  2.76842    
  &  3.67139                &  4.04146                &  4.07319                &  4.07500    
  &  4.07460                &  4.07610    
 \\
$3 \times 10^{9} - 10^{9}$    
  &  2.72615                &  2.64845                &  2.42479                &  2.77894    
  &  3.68546                &  4.05695                &  4.08873                &  4.08978    
  &  4.08135                &  4.07652    
 \\
$3 \times 10^{9} - 10^{8}$    
  &  2.76248                &  2.68807                &  2.47763                &  2.84882    
  &  3.77890                &  4.15986                &  4.19231                &  4.19203    
  &  4.17766                &  4.17398    
 \\
$3 \times 10^{9} - 10^{7}$    
  &  2.77246                &  2.69897                &  2.49216                &  2.86803    
  &  3.80459                &  4.18815                &  4.22083                &  4.22152    
  &  4.22589                &  4.22921    
 \\
$3 \times 10^{9} - 10^{6}$    
  &  2.77348                &  2.70008                &  2.49365                &  2.86999    
  &  3.80721                &  4.19103                &  4.22384                &  4.22638    
  &  4.23167                &  4.23501    
 \\
$3 \times 10^{9} - 10^{5}$    
  &  2.77363                &  2.70025                &  2.49389                &  2.87039    
  &  3.80787                &  4.19223                &  4.22439                &  4.22699    
  &  4.23228                &  4.23561    
 \\
$3 \times 10^{9} - 10^{4}$    
  &  2.77557                &  2.70281                &  2.50008                &  2.89035    
  &  3.85874                &  4.20084                &  4.22507                &  4.22730    
  &  4.23259                &  4.23593    
 \\
$3 \times 10^{9} - 10^{3}$    
  &  2.81121                &  2.75044                &  2.62027                &  3.28700    
  &  4.08933                &  4.23074                &  4.24134                &  4.24270    
  &  4.24796                &  4.25128    
 \\
$3 \times 10^{9} - 10^{2}$    
  &  3.08898                &  3.10503                &  3.45736                &  4.22654    
  &  4.48790                &  4.52946                &  4.53266                &  4.53348    
  &  4.53832                &  4.54137    
 \\
$3 \times 10^{9} - 10$        
  &  3.71314                &  3.95438                &  4.74064                &  5.11335    
  &  5.21097                &  5.22648                &  5.22770                &  5.22826    
  &  5.23210                &  5.23452    
 \\
$3 \times 10^{9} - 1$         
  &  3.99111                &  4.39634                &  5.06123                &  5.33812    
  &  5.40867                &  5.41987                &  5.42076                &  5.42127    
  &  5.42483                &  5.42707    
 \\
\hline
\end{tabular}
\end{table*}

\section{Conclusions}

We have given here details upon the implementation of GS/SOR iterative
processes in 2D cartesian geometry, information which was
unfortunately still missing in the astrophysical litterature. We also
tested, for the first time, such 2D-GS/SOR iterative schemes with a
two-level atom model against original analytical results; a more
comprehensive study, both in 1D and in 2D, is being conducted and
results will be published elsewhere.

Concerning the modelling of illuminated freestanding slabs, even
though we used here a quite simple atomic model, we found it to be a
necessary stage not only to valid our numerical work but also to take
the opportunity to deliver reliable 2D multilevel benchmark results;
typical CPU usage numbers were also given, clearly in favour of the
combination of SOR plus MG methods for complex radiative modelling.

We anticipate that such numerical techniques and benchmark results
will be of interest for the new radiative transfer codes currently in
use or under development, not only for applications in solar physics
but also for interstellar clouds (see e.g., Juvela \& Padoan 2005),
circumstellar environments with winds (see e.g., Georgiev et al. 2006)
or accretion disks (see e.g., Kor$\check{\rm c}$\'akov\'a \& Kub\'at
2005) modelling for instance.

\begin{acknowledgements}
Our warmest thanks go to Dr. Bernard Rutily for the original idea and
fruitful discussions upon the analytical test presented here; we also
thank an anonymous referee for her/his valuable comments which helped us
to clarify some technical points.

\end{acknowledgements}


\begin{appendix}
\section{Test case for a 2D code using 1D reference solutions}

We describe a test case for radiative transfer methods in 2D cartesian
geometry with stationary media, using 1D reference solutions, which
are provided using an analytical method. For this purpose, the ARTY
code is the numerical implementation, whose accuracy is better than
$10^{-10}$, of exact analytical solutions, based on a mathematical
method using the finite Laplace transform (Chevallier \& Rutily 2005,
Chevallier et al. 2003 and references therein).

Our radiative model describes a 2D medium which can scatter in 3D and
is infinite and homogeneous along the $x$-axis ($-\infty \leq x \leq
+\infty$, $0\leq y \leq y_{max}$, $0\leq z \leq z_{max}$), thus
quantities involved in the radiative transfer equation (RTE) do not
depend on $x$. This medium is considered such that there is no
incoming flux on its boundaries along the $y$- and $z$-axes. In order
to compare this 2D case to 1D solutions from ARTY, we consider here
the 2D primary source to be an infinite line along the $x$-axis,
located at the center of the slab, emitting isotropically, and the
medium homogenous and isotropically scattering; the later is also
monochromatic i.e., the RTE does not depend on the frequency (which
will not be mentioned hereafter) as this is the case when we describe
the continuum or a spectral line with the Milne profile, which is
constant over any finite energy range and 0 elsewhere.

We write hereafter the RTE in 2D cartesian geometry, and we show how
to compare this 2D solution integrated on the $y$-axis to a 1D
solution. Table~\ref{tabletest} resumes some values of the 1D solution
at the surface $z=0$. The RTE for our 2D model is (cf. Chandrasekhar
1950, Chap. I, Eq.~(48) or Pomraning 1973, Eq.~(2.60), without
derivative over $x$ though)

\begin{eqnarray}
\label{eq:rte2d}
\lefteqn{ \sin\theta \sin\varphi \frac{\partial I}{\partial y}(y,z,\theta,\varphi)+ \cos\theta \frac{\partial I}{\partial z}(y,z,\theta,\varphi) = } \nonumber \\
&& - \chi [I(y,z,\theta,\varphi)-S(y,z)] \,,
\end{eqnarray}
where $I$ is the specific intensity of the radiative field at $(y,z)$
and in the direction $(\theta,\varphi)$ of the unit vector ${\bf n}$
whose coordinates along $x$, $y$ and $z$ are $\sin \theta \cos
\varphi$, $\sin \theta \sin \varphi$ and $\cos \theta$,
respectively. $\chi$ is the constant opacity of the \emph{homogeneous}
medium, and $S$ is the unknown source function which can be written
\begin{equation}
S(y,z)=S^*(y,z)+ \varpi J(y,z),
\end{equation}
where $S^*$ describes the \emph{primary source function} i.e., the
direct \emph{known} radiative field emitted by internal sources,
$\varpi=(1-\varepsilon)$ is the constant scattering coefficient of the
\emph{homogeneous} medium for simple scattering processes, usually
called albedo, and $J$ is the mean intensity of the radiative field
defined as
\begin{equation}
J(y,z) = \frac{1}{4 \pi} \int_0^{\pi} d\theta \int_0^{2\pi} d\varphi \sin\theta \, I(y,z,\theta,\varphi) \, ;
\end{equation}
the primary source function is
\begin{equation}
S^*(y,z)=\frac{L}{\chi} \,\delta(y-{y_{max} \over 2})
\delta(z-{z_{max} \over 2}) \,,
\end{equation}
where $L$ is the luminosity per unit length along the
$x$-axis. Dividing by $\chi$ means that the source function is an
emissivity divided by the opacity.  In order to use 1D solutions as a
reference, we must integrate the 2D solutions on $y$ over
$[0,y_{max}]$ and on $\varphi$ over $[0,2\pi]$. We thus define new
functions as
\begin{equation}
\tilde{I}(z,\theta) = \frac{1}{2\pi} \int_0^{y_{max}} dy \int_0^{2\pi} d\varphi I(y,z,\theta,\varphi)\, .
\end{equation}
Similarly we define $\tilde{S}(z) = \tilde{S}^*(z) + \varpi
\tilde{J}(z)$, $\tilde{S}^*(z) = L/\chi \,\delta(z-z_{max/2})$,
$\tilde{J}(z)$ and the two successive moments, the radiative flux
$\tilde{H}(z)$ and the radiative pressure $\tilde{K}(z)$ as
\begin{equation}
[\tilde{J}, \tilde{H}, \tilde{K}](z) = \frac{1}{2} \int_0^\pi \tilde{I}(z,\theta) [1,\cos\theta,\cos^2\theta] \sin\theta d \theta \, .
\end{equation}
Integrating over $y$ and $\varphi$, and using the symmetry property
valid for $\varphi \in\, [0,\pi]$: $I(y_{max},z,\theta,\varphi) =
I(0,z,\theta,\pi+\varphi)$, due to the central primary source,
Eq.~(\ref{eq:rte2d}) becomes
\begin{eqnarray}
\label{eq:rte2d1d}
\lefteqn{ \frac{\sin\theta}{\pi} \int_0^{\pi} \sin\varphi I(y_{max},z,\theta,\varphi) d\varphi + \cos\theta \frac{\partial \tilde{I}}{\partial z}(z,\theta) =} \nonumber \\
&&    -\chi [\tilde{I}(z,\theta)-\tilde{S}(z)] \, ,
\end{eqnarray}
where the integral is nul only for $\theta = 0$ or $\pi$; note that
this simplification is fictitious as, even for these angles, the
source function depends on the mean intensity which depends on the
boundaries due to the angular integration. This problem is not
classical and we need to let $y_{max} \rightarrow +\infty$ in order
to suppress this term i.e., the radiation of the primary source is nul
at the infinite and Eq.~(\ref{eq:rte2d1d}) then reduces to the
well-known 1D equation:
\begin{equation}
\label{eq:rte1d}
\mu \frac{\partial I}{\partial z}(z,\mu)=-\chi [I(z,\mu)-S(z)] \, ,
\end{equation}
where $\mu=\cos\theta$.


Equation~(\ref{eq:rte1d}) is usually expressed in optical depth
coordinates $\tau(z)=\int_z^{z_{max}} \chi(z') dz' =
\chi(z_{max}-z)$. We do not write the RTE, but the primary source
function becomes $S^*(\tau)= L \delta(\tau-\tau_{max}/2)$ due to the
Dirac transformation $\delta(z) = \chi \delta(\chi z)$. Accordingly
our 2D primary source function becomes

\begin{equation}
S^*(\tau_y,\tau_z)= \chi \, L
\,\delta(\tau_y-{{\tau_{y_{max}}} \over 2})
\delta(\tau_z- {{\tau_{z_{max}}} \over 2}) \, ,
\end{equation}
where $\tau_y = \chi(y_{max}-y)$ and $\tau_z = \chi(z_{max}-z)$.  In
order to simplify the test of a 2D code with a 1D reference solution,
the values $L=1$ and $\chi=1$ should be used.

We give in Table~\ref{tabletest} some values of the 1D solution at the
surface $z=0$, for the source function, the specific intensity for the
directions of the angular grid used in this paper, and its three first
moments.  When integrating all angles over the azimuthal angle
$\varphi$, the 10-points per octant angular quadrature resume to a
4-points per quadrant, i.e. $[J,H,K](z) = \sum_{i=1,4} w_i
[1,\mu_i,\mu_i^2] I(z,\mu_i)$ for such a case where there is no
incoming flux.  The four directions $\mu_i$ are 0.95118969679,
0.78679579496, 0.57735025883, 0.21821789443 and the integration
weights $w_i$ are 0.063490696251, 0.091383516788, 0.12676086649,
0.21836490929, respectively.  It is interesting to note that, using
the reference solutions, the angular quadrature for J, H and K will
lead to a relative error equal to 0.8\%, 0.3\% and 0.4\% respectively.

\begin{table}
\caption{Reference solutions from the ARTY code at the surface $z=0$
for our test case with $L=1$, $\chi=1$,
$z_{max}=100$ and $\varepsilon=0.01$ (see the text for the values of
$\mu_i$).}
\label{tabletest}
\centering
\begin{tabular}{cc}
\hline \hline
& ARTY results \\
\hline
$S(0)$ & $2.710704655 \times 10^{-4}$ \\
$J(0)$  & $2.738085511 \times 10^{-4}$ \\
$H(0)$ & $1.600980711 \times 10^{-4}$ \\
$K(0)$ & $1.130399095 \times 10^{-4}$ \\
$I(0,\mu_1)$ & $7.837047273 \times 10^{-4}$ \\
$I(0,\mu_2)$ & $6.965481933 \times 10^{-4}$ \\
$I(0,\mu_3)$ & $5.880635905 \times 10^{-4}$ \\
$I(0,\mu_4)$ & $4.026985767 \times 10^{-4}$ \\
\hline
\end{tabular}
\end{table}


\end{appendix}

\end{document}